\documentclass[%
 reprint,
nofootinbib,
 amsmath,amssymb,
 aps,
]{revtex4-1}
\usepackage{booktabs,datetime}
\usepackage{amsmath,amsfonts,amssymb,amsthm,epsfig,epstopdf}
\usepackage{hyperref,physics}
\usepackage[bottom]{footmisc}
\usepackage{mathtools}
\usepackage{makecell}
\usepackage{graphicx}% Include figure files
\usepackage{dcolumn}% Align table columns on decimal point
\usepackage{bm}% bold math
\usepackage{amsthm}
\theoremstyle{plain}

\newtheorem*{theorem*}{Theorem}

\theoremstyle{plain} % just in case the style had changed
\newcommand{\thistheoremname}{}
\newtheorem*{genericthm*}{\thistheoremname}
\newenvironment{namedthm*}[1]
  {\renewcommand{\thistheoremname}{#1}%
   \begin{genericthm*}}
  {\end{genericthm*}}

\begin{document}

\preprint{Thesis, Department of Physics}

\title{Quantum voting and its physical interpretation}% Force line breaks with \\
\thanks{This letter is based on a undergraduate thesis of the author\cite{yoo}}
\author{Woong-seon Yoo}
\email{woongseon.yoo@gmail.com}
\affiliation{Department of Physics and Astronomy, College of Natural Science, Seoul National University}%
\affiliation{Institute for Quantum Advanced Research Korea (i-QUARK), Seoul 06642 KOREA}
\affiliation{Center for Computational and Data-Intensive Science and Engineering, Skolkovo Institute of Science and Technology, Moscow, RUSSIA}

\date{\today}% It is always \today, today,

\begin{abstract}%no-cloning theorem is extended properly for application in Ray space of dimension highter than 1.
Voting is a game with a no-go theorem. New proofs of Arrow's impossibility theorem are given based on quantum information theory. We show that the Arrowian dictator is equivalent to the perfect cloning circuit. We present \textit{Gedankenexperiment} of voting and Bell-like inequalities of voting. We provide the thermodynamic interpretation of voting with Landauer's principle.
% A new proof of Arrow's impossibility theorem is given on voting Hilbert Space: no-cloning theorem is equivalent to Arrow's impossibility theorem. The large limits of Arrow scenario based on Landauer's principle are calculated which explain non-dictatorship on infinite voters and options. Arrowian Bell-like inequalities are also proposed. As a result, the analogy between game theory and quantum theory is highlighted.
\end{abstract}

\maketitle
%\pagenumbering{gobble}
% \tableofcontents
%\newpage

%\clearpage
%\pagenumbering{arabic}
\setcounter{page}{1}
Some say that `information is physical', but physics is the information. Bell-like inequalities capture the non-local nature of quantum theory through Bell violations. For example, in the CHSH experiment\cite{chsh}, classical correlations are bounded by 2, and quantum correlations are bounded by 2$\sqrt{2}$. The CHSH experiment can be transformed into a CHSH game, a measurement game where players can play better with quantum strategies\cite{chshgame}.

Game is a general concept and a toy model for quantum theory\cite{qprisongame}. However, quantizing games itself is not `interesting' enough as mixed strategies correspond to quantum superposition. Quantum game would only be interesting for physicists if there is some physical meaning for games. Nonlocal game is interesting because it is a generalized game of measurements that is a physical view on optimizations and complexity theory.

Voting is a game on the circuit and a communication complexity problem\cite{votingcomplex}. Quantum contextuality is strongly related to communication complexity\cite{bruknerbell}. Differences between classical and quantum communication scenarios are colloquially described by quantum pseudo-telepathy\cite{pseudotelepathy}. Due to pseudo-telepathy, one can design a quantum communication scenario that no classical user can overcome, such as secured voting circuits\cite{thoughtexp}.

The existence of the dictator is the most critical factor in voting. The dictator of voting is a voter whose ballot decides the outcome regardless of others. The dictator always exists under several assumptions, according to Arrow's impossibility theorem. It is possible to prove Arrow theorem with hypercontractivity on Boolean functions as the voting is a classical circuit\cite{kalai}. Recently, quantum Arrow theorem was proposed to be wrong\cite{bao} that is simply because quantum voting breaks the linear order condition:$\textit{pseudo-telepathy}$.

In this letter, we give a physical interpretation of quantum voting, provided with a new proof of Arrow's impossibility theorem. This is possible because voting is a function on the Hilbert space. We use this definition to find a new thought experiment and Bell-like inequalities on the Hilbert space. We also give a sketch to estimate the dictator.

\subsubsection*{Formulation of Arrow's impossibility theorem}

In voting, each voter cast a ballot, preferences on alternatives, to compute an outcome according to the rule. Let $\mathcal{I}$ be the set of voting individuals and $A$ be the set of alternatives. Individuals vote on linear orders on alternatives $\mathcal{P}(A)$ or ballots, and the total state ${\mathcal{T}}$ is a subset of $\mathcal{P}(A)^{\mathcal{I}}$. A preference relation $a_1pa_2$ is a linear order for classical voting, meaning $a_1\geq_{p}a_2$. Classical voting-rule is the function
$r:{\mathcal{T}}\rightarrow\mathcal{P}(A)$, where $r(p^1,p^2,...,p^{|\mathcal{I}|})=p$ and $ p\in\mathcal{P}(A)$. A voting circuit $R$ is defined as $R(p^{ini},p^1,...,p^{|\mathcal{I}|})=(p^{fin},p^1,...,p^{|\mathcal{I}|})$ where $p^{ini}$ is an arbitrary alternative and $p^{fin}=r(p)$ is the result of the voting.

Now, we define conditions of Arrow's impossibility theorem\cite{abramarrow}. Following conditions define the fairness criterion of voting. The voting-rule is called paretorial(\textsc{P}) if $\forall a_1,a_2\in A, \forall p\in{\mathcal{T}}, (\forall i\in\mathcal{I}, a_1{p^i}a_2)\rightarrow a_1r(\mathcal{I})a_2$, independent of irrelevant alternatives(\textsc{IIA}) if $\forall a_1,a_2\in{A}, \forall p,q\in{\mathcal{T}}, p|_{a_1,a_2}=q|_{a_1,a_2}\rightarrow r(p)|_{a_1,a_2}=r(q)|_{a_1,a_2}$ and unrestricted domain(\textsc{UD}) if $\forall p\in\mathcal{P}(\{a_1,a_2,a_3\})^{\mathcal{I}}.$ $\exists q\in{\mathcal{T}}.$ $q|_{a_1,a_2,a_3}=p$. The voting-rule is called dictatorial if $\exists i\in\mathcal{I}, \forall a_1,a_2\in{A}.\forall{p}\in{\mathcal{T}}. a_1p_ia_2\rightarrow a_1r(p){a_2}$ with the dictator $i$. Fair voting is the voting with conditions above.
\begin{namedthm*}{Arrow's impossibility theorem}[]
If $r:\textsc{UD,P,IIA}$ and $2<|A|,|\mathcal{I}|<\infty$ then $r$ is dictatorial: fair voting always has the dictator.
\end{namedthm*}

Several variations of fair voting exist and two of them are to be discussed in following. First, there exists voting with nonlinear order, $\textit{cardinal voting}$, that Arrow theorem does not hold. Thereby, we design a quantum cardinal voting whose restriction is classical and ordinal. Second, there exists voting with infinite voters and alternatives. Fishburn\cite{fishburn} showed the existence of fair and nondictatorial voting with infinite voters through hyperfilters. We apply Landauer principle to interpret infinite voting.

\subsubsection*{Quantum voting on the Hilbert space}
Quantum voting is a cardinal voting on the Hilbert space with the cardinality defined as the probability of every classical ballot. Casting ballots are analogous to measuring states, hence, it is natural to correspond the voting with the Hilbert space. Quantum voting is the function on following Hilbert space.
\begin{namedthm*}{Ballot Hilbert space}
Ballot Hilbert space $\mathcal{H}_B$ with $n$ alternatives is a finite separable Hilbert space whose dimension is larger than the number of possible ballots, $n!$ and a subset of the basis assigned with ballots.
\end{namedthm*}
Quantum voting rule is defined on the entire Hilbert space such that $\hat{r}:\mathcal{H}^{\otimes m}\rightarrow\mathcal{H}$ and the corresponding quantum circuit with an ancila, $\hat{\mathcal{R}}:\mathcal{H}^{\otimes m+1}\rightarrow\mathcal{H}^{\otimes m+1}$ which is a unitary operator that transforms the ancila into the voting result $\hat{r}(\hat{\mathcal{T}})$. 

Classical voting is a restriction of quantum voting. A (sub)set of the orthogonal basis is the collection of possible ballots or linear orders on alternatives. Classical voting $\hat{r}_c$ and circuit $\hat{\mathcal{R}}_c$ are defined on the finite product of orthogonal rays $\gamma$ of Hilbert space, $\hat{r}_c$: $\gamma(\ket{p})^{\otimes}\rightarrow\gamma(\ket{p})$.

Using Ballot Hilbert space is not a unique method to represent voting. A classical ballot $a_1a_2...a_m$ is composed of $\binom{m}{2}$ Boolean preference relations: $a_1pa_2,a_1pa_3,..$. Similarly, a quantum ballot is composed of $\binom{m}{2}$ preference relations: $\ket{a_1a_2...a_m}=\ket{a_1a_2}\ket{a_1a_3}...$. It is fruitful to study quantum voting with quantum Boolean functions but the main point of this letter use the qudit representation to claim following: the dictatorial circuit is the cloning machine.
\subsubsection*{New proof with no-cloning theorem}
In this section, we show that a dictatorial circuit is a universal cloning operator. In the quantum circuit, a dictatorial circuit is defined as a circuit $\hat{\mathcal{R}}\ket{a}\ket{p^1}...\ket{p^{m}}=\ket{r}\ket{p^1}...\ket{p^{m}}$ such that $\ket{r}=\ket{p^1}$ up to a permutation on the set of voters. For a dictator, $\ket{r}=\ket{p^1}$ should be satisfied regardless of other ballots $\ket{p^2}...$, which is equivalent to the statement of no-cloning theorem.

\begin{namedthm*}{No cloning theorem}
There is no unitary operator $\mathcal{U}$ on $\mathcal{H}\otimes\mathcal{H}$ such that $\mathcal{U}\ket{\psi}\ket{\phi}=e^{ic}\ket{\psi}\ket{\psi}$, where $\ket{\psi},\ket{\phi}$ are normalized states of $\mathcal{H}$, c is some real number.
\end{namedthm*}
Quantum monogamy is due to superpositions between orthogonal states of dimension bigger than 2\cite{oriclon}. Likewise, Arrowian dictator is due to non-linear orders of alternatives larger than 2. We use this to observe that the universal cloning operator is the dictatorial circuit.
\begin{namedthm*}{Theorem}
The voting circuit with the dictator is equivalent to the cloning operator. In other words, there exists the dictator iff preference states with Arrow conditions are orthogonal on VHS.
\end{namedthm*}

The voting circuit with the dictator is the cloning operator. The voting circuit with the dictator is represented as $R_d(p^{ini},p^{dic},...)=(p^{dic},p^{dic},...)$ which is the cloning operator on VHS$\otimes$VHS.

Let's assume that preference states are orthogonal which is the classical voting. Then, there exists the dictator by Arrow's impossibility theorem. Moreover, the voting circuit with the dictator is the cloning operator. Now, let's assume that preference states are not orthogonal which is the quantum voting. By no cloning theorem, the cloning operator does not exist, hence there is no dictator.

Another proof can be stated with the Kochen-Specker set\cite{ks,pseudotelepathy}. A Kochen-Specker set on the Hilbert space is a subset $\mathcal{S}$ of unit vectors such that there is no function $r:\mathcal{S}\rightarrow\{0,1\}$ s.t if $\mathcal{S}$ contains an orthonomal basis $b$, $\sum_{b}r=1$. Classical voting rule with the dictator $\hat{r}_c$ is the function satisfying the property. The function $r$ defines a dictatorial voting-rule $\hat{r}_v$ for $v\in b, r(v)=1$.
\subsubsection*{Thought experiment of voting}
We derive Bell-like inequalities that distinguish quantum voting from classical voting. Bell-like inequalities distinguish quantum systems from classical systems with maximal violations which also mean that systems can be interpreted as $(r,s,t)$ Bell scenario with $r$ observers, $s$ settings and $t$ outcomes. Fair voting with $m$ voters and $n$ alternatives equipped with $m$ Ballot Hilbert spaces $\mathcal{H}_B^{\otimes m}$ or $(m,n)$ Arrow scenario is also a Bell scenario.

Voting is a quantum pseudo-telepathy scenario. It is impossible to evade dictatorship from classical Arrow scenario of fair voting, because voters are not allowed to communicate each other. It is possible to achieve fair voting for quantum voters, though voters are not allowed to communicate each other. As pseudo-telepathy is a feature of nonlocality, the existence of Bell-like inequalities for voting is guaranteed.

There are two types of Bell-like inequalities for voting: trivial and non trivial inequalities. Nontrivial inequalities always include the dictator and trivial inequalities are independent from the dictator. Finding trivial inequalities are equivalent to Bell inequalities of measurements and relatively easy to find.

For instance, let us assign two observers to $(m,3)$ Arrrow scenario: Alice for a voter's outcome and Bob for the voting outcome. One can embed 6 classically possible states into spin $\frac{1}{2}$ variables:
% Voters number of $m$ are assigned with the Ballot Hilbert space of $n$ alternatives, $\mathcal{H}^{\otimes m}_B$. Each voter casts a ballot defined as a measurement. Angles of measurements for voters are identified by results of measurements, linear combinations of alternatives $\sum p^j_i\ket{r_i}$. The configuration without the voting outcome is a $(m,)$ 
% Voting is a generalized Bell scenario. In EPR scenario, Alice and Bob measures two spin $\frac{1}{2}$ entangled particles with each measurement axis. 
% in EPR scenario, represented as $\frac{1}{2}(\mathbb{I}\pl\vec{r}\cdot\vec{\sigma})$, where $|\vec{r}|=1$.
% $(m,3)$-Arrowian scenario is original EPR-Bell scenario. By assigning a ray $\gamma_{a_i}$ to profiles where $a_i$ is at the second place, like
% $\ket{a_{1}a_{2}a_{3}}\rightarrow r^+_2, \ket{a_{3}a_{2}a_{1}}\rightarrow r^-_{2}$ where $r^\pm_{2}\in\gamma_2$.
$\ket{a_{1}a_{2}a_{3}}\rightarrow r^+_2, \ket{a_{3}a_{2}a_{1}}\rightarrow r^-_{2}, \ket{a_{3}a_{1}a_{2}}\rightarrow r^+_{1}, \ket{a_{2}a_{1}a_{3}}\rightarrow r^-_{1}, \ket{a_{2}a_{3}a_{1}}\rightarrow r^+_{3}, \ket{a_{1}a_{3}a_{2}}\rightarrow r^-_{3}$. By embedding ballots into measurement axis, it is possible to derive trivial inequalities of CH-type. Ballots of three alternatives are made of 3 preference relations:$a_1{p}a_2, a_2{p}a_3, a_3pa_1$.

% CH correlation reads $\bra{r_x}(A\cdot{G_x})(B\cdot{G_x})\ket{r_x}=-A\cdot{B}$. This extends to $(m,n)$ Arrow scenario as $C(a,b)\equiv(p^+_a-p^-_a)(p^+_b-p^-_b)=(2p^+_a-1)(2p^+_b-1)=(2\bra{r_x}G_{a}\ket{r_x}-1)(2\bra{r_x}G_{b}\ket{r_x}-1)$. Classical correlation reads $C_c(a,b)=(2\delta_{{r_x},a}-1)(2\delta_{{r_x},b}-1)$ and quantum correlation reads $C(a,b)=(2\bra{r_x}G_{a}\ket{r_x}-1)(2\bra{r_x}G_{b}\ket{r_x}-1)$.
Finding non-trivial inequalities, which include the dictator, are hard to design. Bell inequality distinguishes quantum measurements because quantum measurements can exceed the classical Bell bound. However, the dictatorship does not mean the Bell bound. In Bell scenario, some quantum measurements do not violate the classical Bell bound, whereas there is no dictatorial quantum voting. Therefore, the non-trivial Bell-like inequality should be in a combinatorial form. The non-trivial Bell-like inequality will fully answer to the question of Abramsky to relate nonlocality to $\textsc{IIA}$\cite{abramarrow}.

\subsubsection*{Thermodynamic interpretation of dictator}
No-go theorems state that classical information can be cloned and erased, whereas quantum information cannot be cloned and erased. Plenio\cite{forget} showed that the cloning quantum information violates Landauer principle.
\begin{namedthm*}{Landauer's principle}
Erasing a classical bit costs energy at least $kTlog2$
\end{namedthm*}
In their arguments, a cloning quantum information leads to the increased entropy, hence one gets information lager than its source which is violation of the thermodynamic law. We apply the argument to the voting as the dictatorial circuit is the perfect cloning circuit.

Due to changes in the information degree of freedom, Landauer limit should be $kTlogD$, instead of $kTlog2$. In classical voting, there always exists the dictator, thereby every fair voting rule is dictatorial. Let Alice know the dictator and the ballot and let Bob guess the ballot of the dictator without any information. Naive Bob tries $a$ $priori$ strategy that is a series of trials-and-errors followed by erasures. Bob need to guess which voter is the dictator and the dictatorial ballot with and without remembering his own guesses.

If Bob can remember his guesses, he only needs to erase his guesses finitely. For finding the dictator, it would take $kTlogn!$, and for finding his ballot, it would take $kTlog(m!)!$. If Bob cannot remember his guesses, he needs to erase his guesses infinitely which is $(n-1)kTlogn$ and $(m!-1)kTlogm!$. Let us denote the total expected energy $\mathcal{E}$ as sum of expected energy for the dictator $\mathcal{E}_1$ and expected energy for the ballot $\mathcal{E}_2$.

Let us consider single particle cases of voter and alternative. When there is only a voter($n=1$), the voter is the dictator by the definition, therefore no energy is required to find the dictator. Same holds for the single alternative case, as well. Zero energy of finding means Bob can always find operator with the $a$ $priori$ strategy. Even for the finite energy, there exist unitary circuits corresponds to such energy.

There are two infinite limits of voters and alternatives. First, if the number of voters goes to infinity, Arrow theorem does not hold\cite{fishburn}. According to thermodynamic interpretation of voting, $\mathcal{E}_1$ goes to infinity as $n$ goes to infinity. There is no corresponding unitary circuit corresponds to infinite energy and Arrow theorem does not hold. Second limit is the infinite alternative which is the quantum voting. There are infinite combinations of orthogonal ballots with superpositions in quantum voting. 

This letter is based on Yoo's undergraduate thesis submitted to Prof. Soo-Jong Rey, Department of Physics, Seoul National University. The author enjoyed valuable discussions with Kabgyun, Jinhyoung, Mauro, Dax, Mason, Nathan, Roman and Zoltan.
%\frenchspacing
\providecommand{\bysame}{\leavevmode\hbox to3em{\hrulefill}\thinspace}
\providecommand{\MR}{\relax\ifhmode\unskip\space\fi MR }
\providecommand{\MRhref}[2]{%
  \href{http://www.ams.org/mathscinet-getitem?mr=#1}{#2}
}
\providecommand{\href}[2]{#2}
\bibliographystyle{unsrt}
% \nocite{*}
\bibliography{reference}

\begin{thebibliography}{10}

\bibitem{yoo}
Woongseon Yoo.
\newblock No cloning proof and bell-like inequality of arrow's impossibility
  theorem.
\newblock {\em Undergraduate Thesis, Seoul National University}, June, 2018.

\bibitem{chsh}
John~F Clauser, Michael~A Horne, Abner Shimony, and Richard~A Holt.
\newblock Proposed experiment to test local hidden-variable theories.
\newblock {\em Physical review letters}, 23(15):880, 1969.

\bibitem{chshgame}
Anna Pappa, Niraj Kumar, Thomas Lawson, Miklos Santha, Shengyu Zhang, Eleni
  Diamanti, and Iordanis Kerenidis.
\newblock Nonlocality and conflicting interest games.
\newblock {\em Physical review letters}, 114(2):020401, 2015.

\bibitem{qprisongame}
Jens Eisert, Martin Wilkens, and Maciej Lewenstein.
\newblock Quantum games and quantum strategies.
\newblock {\em Phys. Rev. Lett.}, 83:3077--3080, Oct 1999.

\bibitem{votingcomplex}
Vincent Conitzer and Tuomas Sandholm.
\newblock Communication complexity of common voting rules.
\newblock In {\em Proceedings of the 6th ACM conference on Electronic
  commerce}, pages 78--87. ACM, 2005.

\bibitem{bruknerbell}
\ifmmode~\check{C}\else\v{C}\fi{}aslav Brukner, Marek \ifmmode~\dot{Z}\else
  \.{Z}\fi{}ukowski, Jian-Wei Pan, and Anton Zeilinger.
\newblock Bell's inequalities and quantum communication complexity.
\newblock {\em Phys. Rev. Lett.}, 92:127901, Mar 2004.

\bibitem{pseudotelepathy}
Gilles Brassard, Anne Broadbent, and Alain Tapp.
\newblock Quantum pseudo-telepathy.
\newblock {\em Foundations of Physics}, 35(11):1877--1907, 2005.

\bibitem{thoughtexp}
Marianna Bonanome, Vladim\'{\i}r Bu\ifmmode~\check{z}\else \v{z}\fi{}ek, Mark
  Hillery, and M\'ario Ziman.
\newblock Toward protocols for quantum-ensured privacy and secure voting.
\newblock {\em Phys. Rev. A}, 84:022331.

\bibitem{kalai}
Gil Kalai.
\newblock A fourier-theoretic perspective on the condorcet paradox and arrow's
  theorem.
\newblock {\em Advances in Applied Mathematics}, 29(3):412 -- 426, 2002.

\bibitem{bao}
N~Bao and N~Yunger~Halpern.
\newblock Quantum voting and violation of arrow's impossibility theorem.
\newblock {\em Phys. Rev. A}, 95:062306.

\bibitem{abramarrow}
Samson Abramsky.
\newblock Arrow's theorem by arrow theory.
\newblock 2014.

\bibitem{fishburn}
Peter~C Fishburn.
\newblock Arrow's impossibility theorem: Concise proof and infinite voters.
\newblock {\em Journal of Economic Theory}, 2(1):103 -- 106, 1970.

\bibitem{oriclon}
W.~K Wootters and W.~H Zurek.
\newblock A single quantum cannot be cloned.
\newblock {\em Nature}, 299(5886):802--803, 1982.

\bibitem{ks}
Simon Kochen and Ernst~P Specker.
\newblock The problem of hidden variables in quantum mechanics.
\newblock In {\em The logico-algebraic approach to quantum mechanics}, pages
  293--328. 1975.

\bibitem{forget}
M.~B. Plenio and V.~Vitelli.
\newblock The physics of forgetting: Landauer's erasure principle and
  information theory.
\newblock {\em Contemporary Physics}, 42(1):25--60, 2001.

\end{thebibliography}
\end{document}